\documentclass[12pt]{elsarticle}

\usepackage{amsfonts}
\usepackage{amssymb}
\usepackage{graphicx}
\usepackage{color}
\renewcommand{\L}{\left}
\newcommand{\R}{\right}

\newcommand{\ex}[2][-\imath]{e^{#1 #2}}
\newcommand{\ii}{\imath}
\newcommand{\shalf}{\frac{1}{\sqrt{2}}}
\newcommand{\half}{\frac{1}{2}}

\newcommand{\f}[1]{\textrm{#1}}
\newcommand{\B}[1]{\mathbf{#1}}

\newcommand{\eqa}[1]{ \begin{eqnarray}#1\end{eqnarray}}
\renewcommand{\vector}[4][r]{
\left(\begin{array}{#1}
#2 \\
#3 \\
#4 \\
\end{array}\right)}

\begin{document}

\title{An extended representation of three spin component Bose Einstein Condensate solitons}

\author{Piotr Szankowski\footnote{Corresponding author: e-mail: piotr.szankowski@fuw.edu.pl}, Marek Trippenbach}

\address{Institute of Theoretical Physics, University of Warsaw, ul. Ho\.{z}a 69, PL--00--681 Warszawa, Poland. }

\author{Eryk Infeld}

\address{National Centre for Nuclear Research, ul. Ho\.{z}a 69, PL-00-681 Warsaw, Poland}

\begin{abstract}
We consider a three spin component Bose Einstein Condensate as described by as many coupled nonlinear Schroedinger equations. For a very special ratio of the coupling constants, exact $N$ soliton solutions to this set of equations are known. Here we find a simple representation including the $N=1$ solution based on the symmetry of the equations.  This symmetry is described by a linear operator, the nonlinearity of NLS notwithstanding. Our useful representation opens the door to the nonintegrable case of {\it general} coupling constants. A new class of solutions is found.
\end{abstract}

\maketitle

Bose Einstein condensates (BEC) offer an unusual opportunity to test quantum effects on the mesoscopic scale. In particular spinor BEC condensates furnish an excellent example of this possibility \cite{Shenoy+Ho:BinaryMixturesOfBEC,Machida:BECWithInternalDoF,Klausen:SpinorBECinRb}. In 1998, the MIT group published observations of the first example of a spinor condensate, in which they transferred a spin-polarized $^{23}$Na condensate, created in a traditional magnetic trap, into a dipole trap formed by the focus of a far-off-resonant laser \cite{Ketterle:OpticallyTrapedBEC}.
Soon after that they observed the formation of spin domains resulting from population exchanges within the $F = 1$ manifold via spin-relaxation collisions. These collisions allowed for a redistribution of the spin population according to the constraints of the local magnetic field \cite{Ketterle:SpinDomainsinSpinorBEC,Kronjager:TunedSpinDynamics,Kronjager:EvolutionOfSpinorBEC,Chapman:NatureSpinorBEC}.
Multiple condensates have numerous possible applications, such as high resolution interferometry \cite{Vengalattore:MagnetometrySpinorBEC}, multi channel signals and their switching, and even optodynamical devices \cite{Stamper-Kurn:OptodynamicsSpinorBEC}

Here consider a spinor condensate formed by atoms with spin F = 1, described by a three components microscopic wave function. Some very special multi-component vector solitons have already been predicted \cite{Wadati:SpinorSolitons,Wadati:BrightSpinorSolitons,Wadati:DarkSolitons}. Their stability has been considered extensively, see for example \cite{Doktorov:PRA2008,Li+Liu+Malomed:PRA2005,eimt:citup,mtei:citup}. In our previous work we investigated polar and ferromagnetic solitons, introduced in \cite{Wadati:SpinorSolitons,Wadati:BrightSpinorSolitons,Wadati:DarkSolitons}. We showed that they can be generalized for a wider range of coupling parameters. By studying their collisions we also discovered a new class of solitary waves, which we called oscillatons. These include the polar and ferromagnetic solitons \cite{Szankowski:PRA,Szankowski:10}.  In further studies we recognized that all bright solitons and solitary waves in spin F = 1 systems are oscillatons! This is discussed in the present work.  Even though the collisions of oscillatons are inelastic, the emerging species are once again oscillatos \cite{Szankowski:PRA,Szankowski:10}. They form an "invariant subdivision".


We consider a one dimensional, dilute gas of $N$ bosonic atoms with hyperfine spin $F=1$ at very low temperatures. Interactions between atoms can be parametrized by two coupling constants $c_0=4\pi\hbar^2 (a_0+2a_2)/3M$ and $c_2=4\pi\hbar^2(a_0-a_2)/3M$, where the $a_f$ are s-wave scattering lengths of total spin $f$ channels.
 The three component, macroscopic wave function,
$\Phi_m(x,t)$ with $m= -1,0,1$, satisfies the Gross - Pitaevskii equation, which in dimensionless units has the form
\eqa{\label{eq:MainEQ}
\imath\partial_t\Phi_m & = &
-\L(\partial_x^2 + \sum_{m'}|\Phi_{m'}|^2\R)\Phi_m \\ \nonumber
& & -\gamma\sum_\alpha\L(\sum_{m',\,m''}\Phi^\ast_{m'}f_{m',\,m''}^\alpha\Phi_{m''}\R)
\sum_{m'}f_{m,\,m'}^\alpha\Phi_{m'}.
}
Here $\hat f^\alpha \;(\alpha=1,2,3)$ are the angular momentum operators in a 3x3 representation and $\gamma=-c_2/|c_0|$
\footnote{If we choose as our length scale $x\rightarrow \frac{\hbar^2\,L^2}{N|c_0|}\,x$, and time scale $t\rightarrow \frac{2M}{\hbar}\left(\frac{\hbar^2\,L^2}{N|c_0|}\right)^2$ ($N$ is the number of particles and $L$ is some characteristic length for given system, e.g. the transverse size of the trap confining a quasi one - dimensional condensate), all coefficients but the ratio between self and cross nonlinear coupling, $-c_2/|c_0|$, which we denote by $\gamma$, will be equal to $\pm 1$.}

This equation is symmetric under Galilean boosts and translations, $\B\Phi ' (x,t) =\hat{\mathcal B}(p,\delta x)\B\Phi =\ex[\ii]{(p (x-\delta x)-p^2 t)}\B\Phi(x-\delta x-2pt,t)$ and also under rotations,
$\B\Phi ' = \ex{\tau}\hat{\mathcal U}(\phi,\theta,\psi)\B\Phi = \ex{\tau}\ex{\phi\hat f^3}\ex{\theta\hat f^2}\ex{\psi\hat f^3}\B\Phi$. Here, the operation $\hat{\mathcal U}$ represents rotations of the quantization axis. In physical terms, $\hat{\mathcal U}$ corresponds to turning a Stern - Gerlach type apparatus used to measure the wave function components. We will put this symmetry to work.

Ieda et al considered the case $c_0 = c_2 = c<0$, equivalent to $\gamma =1$ \cite{Wadati:SpinorSolitons,Wadati:BrightSpinorSolitons,Wadati:DarkSolitons}. Equation (\ref{eq:MainEQ}) then is completely integrable. Bright $N$-soliton solutions where found. For this case, binary collisions of solitons are illustrated and extremely long formulas describing them are given in \cite{Wadati:SpinorSolitons,Wadati:BrightSpinorSolitons,Wadati:DarkSolitons}. The general {\it one-soliton} solution is given by
\eqa{\label{eq:WadatiOneSoliton}
\B\Phi_{sol} = \frac{2\sqrt\mu\,e^{- \imath \mu t}}{1+4\mu e^{-2\sqrt\mu(x-x_0)}+\frac{1}{16\mu}|\langle\hat T\rangle|^2e^{2\sqrt\mu(x-x_0)}} \cdot & \nonumber\\
\left[2\sqrt\mu\, e^{-\sqrt\mu(x-x_0)}\;\chi+\frac{1}{4\sqrt\mu}\langle\hat T\rangle^\ast e^{\sqrt\mu(x-x_0)}\;\hat T\chi\right] &  ,
}
where $x_0 =- \ln( 8\mu/|\langle\hat T\rangle|)/2$, $\chi$ is a normalized polarization spinor $\chi=(\chi_1,\chi_0,\chi_{-1})$ ($\chi^\dagger\chi = \sum_m|\chi_m|^2=1$) and $\hat T$ is a time reversing operator, defined as
$\hat T\chi = ( \chi_{-1}^\ast,-\chi_0^\ast,\chi_{+1}^\ast )^T$.
 The quantity $\langle\hat T\rangle$ is a mean value of this operator: $\chi^\dagger \hat T\chi$. Our first goal was to find an expression for one-soliton solution, which is more transparent and expressed in terms of recognizable physical quantities. We arrived at:
\begin{equation}\label{eq:SplitOscillaton1}
\B\Phi =
e^{-\imath\tau}\hat{\mathcal U}(\phi,\theta,\psi)
\vector[c]
{ \sqrt{\mu_+}\;\f{sech}\left[\sqrt{\mu_+}\left(x + \Delta x\right)\right]\ex[\ii]{\mu_+ t}}
{0}
{  - \sqrt{\mu_-}\;\f{sech}\left[\sqrt{\mu_-}\left(x - \Delta x\right)\right]\ex[\ii]{\mu_- t}}.
\end{equation}
Aside from rotation, this expression appears to be a superposition of two oppositely polarized displaced solitons. It reduces to Ieda's one soliton formula Eq.~(\ref{eq:WadatiOneSoliton}) when $\mu_+=\mu_- = \mu$ and $\Delta x$ given by:
\begin{equation}
\Delta x = \frac{1}{2\sqrt\mu} \ln\left( \frac{1+\sqrt{1-|\langle \hat T\rangle|^2}}{|\langle\hat T\rangle|} \right).
\end{equation}
For a given value of $\langle\hat T\rangle$ a direct relation between the angles of rotation and the components of the polarization spinor used in the papers of Ieda {\it et al} \cite{Wadati:SpinorSolitons,Wadati:BrightSpinorSolitons} is
\begin{eqnarray}
\tau &=&\frac{\pi}{2}-\half \arg\langle\hat T\rangle,\nonumber\\
\beta &=&\sqrt\mu\Delta x -\ii\psi,\nonumber\\
\chi_0 &=& \ex{\tau}\sqrt{|\langle\hat T\rangle|}\sin\theta\cosh\beta,\nonumber\\
\chi_{+1}  &=& \shalf\ex{(\tau+\phi)}\sqrt{|\langle\hat T\rangle|}\left[\cos^2\left(\frac{\theta}{2}\right)\,e^{\beta}
-\sin^2\left(\frac{\theta}{2}\right)\,e^{-\beta}
\right],\nonumber\\
\chi_{-1} &=&\shalf\ex{(\tau-\phi)}\sqrt{|\langle\hat T\rangle|}\left[\sin^2\left(\frac{\theta}{2}\right)\,e^{\beta}
-\cos^2\left(\frac{\theta}{2}\right)\,e^{-\beta}
\right].\nonumber
\end{eqnarray}

\begin{figure}[tbp]
\includegraphics[scale=0.75]{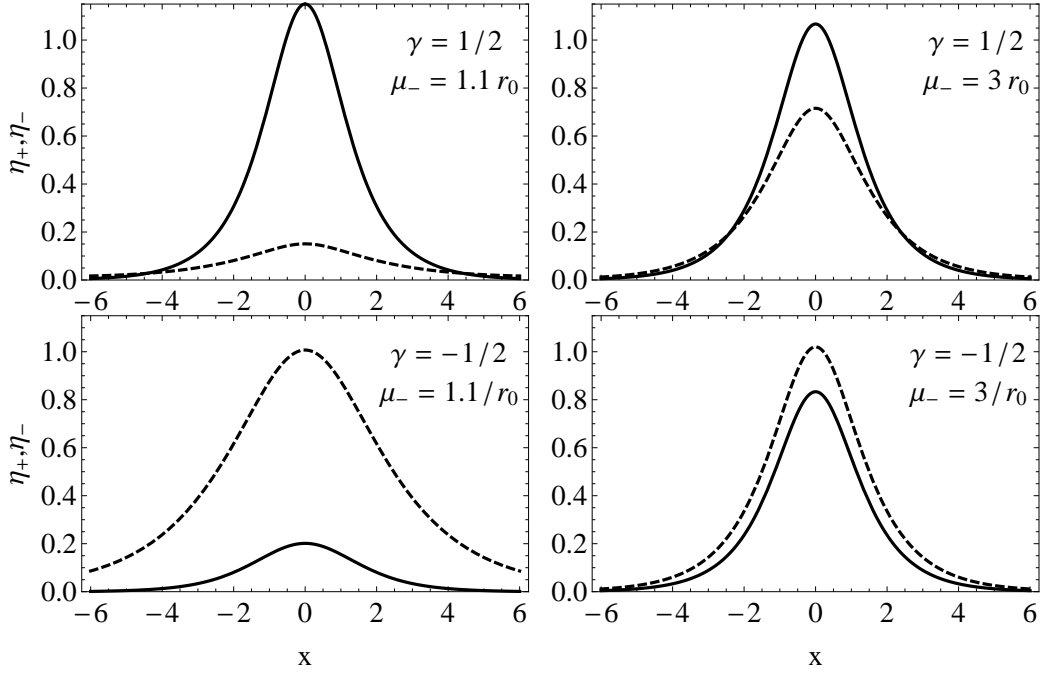}
\caption{Four examples of solutions to Eq.~(\ref{eq:OscillatonEq})  for
oscillaton components $\eta_+$ (continuous line) and $\eta_-$ (dashed line), $\gamma = \pm 1/2$. }\label{fig1}
\end{figure}

For $|\langle \hat T\rangle| \to 1$ and $|\langle \hat T\rangle| \to 0$ single soliton Eq.~(\ref{eq:WadatiOneSoliton}) reduces to polar and ferromagnetic solitons\footnote{In case of $\langle \hat T\rangle \to 0$ the displacement $\Delta x$ tends to infinity and the spinor components become independent. Each is a ferromagnetic soliton.} respectively. In these limits, even without our improvements Eq.~(\ref{eq:SplitOscillaton1}) is recovered. In previous papers we only studied those two cases. The whole spectrum of one soliton solutions with $0<|\langle \hat T\rangle|< 1$ seemed to be fundamentally different then the above limiting cases and solutions of the form (\ref{eq:SplitOscillaton1}). This apparent distinction is now abolished.

When $\Delta x$ is arbitrary and $\mu_+$ is not equal to $\mu_-$ our Eq.~(\ref{eq:SplitOscillaton1}) takes us out of Ieda's one soliton solutions as given by Eq.~(\ref{eq:WadatiOneSoliton}). To recreate this more general form in Ieda's terms, one would have to go to their two soliton solutions \cite{Wadati:SpinorSolitons,Wadati:BrightSpinorSolitons}.

Most importantly the form of Eq.~(\ref{eq:SplitOscillaton1}) allows for an easy generalization to a much broader class of solutions when $\gamma \neq 1$. This is a central result of our Letter. Its general form is
\begin{equation}\label{eq:SplitOscillaton}
\B\Phi =
\ex{\tau}\hat{\mathcal U}(\phi,\theta,\psi)\vector[c]
{ \eta_+(x)\ex[\ii]{\mu_+ t}}
{0}
{  - \eta_-(x)\ex[\ii]{\mu_- t}}.
\end{equation}
For $\gamma \neq 1$ the $\eta$'s are in general no longer simple $\f{sech}$ functions. Substituting (\ref{eq:SplitOscillaton}) into Eq.~(\ref{eq:MainEQ}) yields the following system of two coupled \emph{ordinary} differential equations, which the $\eta$'s must satisfy (the prime denotes $x$ differentiation):
\begin{equation}\label{eq:OscillatonEq}
-\mu_\pm + (1+\gamma)\eta_\pm^2 + (1-\gamma)\eta_\mp^2 + \frac{\eta_\pm ''}{\eta_\pm} =0.
\end{equation}
For each value of $\gamma$ there is a limit on possible values of the ratio of $\mu_-/\mu_+$. The range of this ratio was found in \cite{Yang:ClassificationOscillatons}
\begin{equation}
\left\{\begin{array}{lll}
r_0\,\leq\,\frac{\mu_+}{\mu_-}\,\leq\,\frac{1}{r_0} &\f{, for}&\gamma>0\\
\frac{1}{r_0}\,\leq\,\frac{\mu_+}{\mu_-}\,\leq\,r_0 &\f{, for}&\gamma<0\\
\end{array}\right.
\end{equation}
where
\begin{equation}\label{eq:mu0}
r_0=\frac{1}{4}\left(\sqrt{1+8\frac{1-\gamma}{1+\gamma}}-1\right)^2
\end{equation}
In the limit $\gamma=1$, Eqs.~(\ref{eq:OscillatonEq}) separate and the solutions are $\f{sech}$ functions with arbitrary shifts and amplitudes, as mentioned above. In Fig.~(\ref{fig1}) we present four examples of $\eta$ pairs, for two different values of $\gamma = \pm 1/2$ and zero shifts.

In a numerical study we took a shifted solution given by (\ref{eq:SplitOscillaton}) and slowly changed $\gamma$ starting with one. Even for small changes of $\gamma$ the components $\eta_+$ and $\eta_-$ attract or repel each other, due to the nonlinear coupling, and they begin to move with respect to each other. On top of that, we also observed inevitable excitations and radiation. However, if there are no shifts, an adiabatic change of $\gamma$ leads to adiabatic deformation of the $\eta$'s and $\mu$'s, preserving the form of Eq.~(\ref{eq:SplitOscillaton}).

\begin{figure}[tbp]
\centering
\includegraphics[scale=0.5]{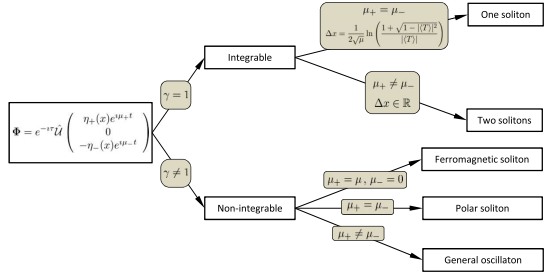}
\caption{Schematic diagram of cases considered in this Letter.}\label{fig2}
\end{figure}

Equation (\ref{eq:OscillatonEq}) is considered in the literature in the context of solitary waves in a two component system \cite{Yang:ClassificationOscillatons}.
However, this is a merely formal similarity. In \cite{Yang:ClassificationOscillatons} the author considers a two component system, not supporting the possibility of a transfer between components.
Here we focus on the $F=1$ manifold of a spinor condensate and this is a three component system. In our case the operator $\hat{\mathcal U}$ leads not only to the usual shift in population, but also to population oscillations between components. They are possible due to the existence of source terms in the Eq.~(\ref{eq:MainEQ}), proportional to $\gamma$. Due to these oscillations we call these entities {\it oscillatons} (The name is not entirely new - oscillatons appear as solutions to the Einsten-Klein-Gordon equations, see E. Seidel et al, Phys. Rev. Lett. {\bf 66}, 1659 (1991) and {\bf 72}, 2516, (1994)). The explicit time dependence of the density of the three components of the oscillatons are given by
\begin{eqnarray}
|\Phi_0|^2 &=& \half\sin^2\theta\left(\eta_+^2+\eta_-^2
  +2\eta_+\eta_-\cos\left[\omega\,t-2\psi\right]\right),\\
|\Phi_{\pm 1}|^2 &=& \eta_\pm^2\sin^4\frac{\theta}{2}+\eta_\mp^2\cos^4\frac{\theta}{2}
 -\half\eta_+\eta_-\sin^2\theta\cos\left[\omega\,t-2\psi\right],
\end{eqnarray}
which exhibit oscillatory behavior with frequency $\omega=\mu_+-\mu_-$. Fig.~\ref{fig2} places the Ieda {\it at al} solution in a wider context of our family of calculations. For more details of oscillaton collisions consult our previous papers \cite{Szankowski:10,Szankowski:PRA}.

Another case of interest is when $\gamma = 0$. In many experimental applications $\gamma$ is indeed very small (it is proportional to the difference between the two characteristic scattering lengths $a_0$ and $a_2$). Take for example ${ }^{87} \f{Rb}$ with $\gamma\sim 0.01$ \cite{Klausen:SpinorBECinRb}. When $\gamma$ is exactly zero, our system is once again integrable. There is no transfer between components of the wavefunction and the rotation $\hat{\mathcal U}$ becomes a trivial operation. Notice that now $r_0$ must be exactly one and hence $\mu_+ = \mu_-$. There is no room for oscillations since the frequency $\omega = \mu_+ - \mu_-$ is zero. This is not surprising, since for $\gamma = 0$ the equation of motion no longer admits spin mixing.

In conclusion, when treating spinor Gross-Pitaevskii equations ($F=1$), it has been possible to improve on the Ieda {\it et. al.} formula for a one soliton solution. The adopted form, apart from being simpler and easier to manipulate, covers more cases and leads us into the world of \emph{oscillatons}, entities with interesting properties including population oscillations between spin components following from the rotational symmetry of the system. These entities have been found recently \cite{Szankowski:PRA,Szankowski:10}. Now they follow from a systematic procedure.

This project was financed by the National Science Centre. P. S. acknowledges financial support by the Foundation for Polish Science International PhD Projects Programme co-financed by the EU European Regional Development Fund.

\section*{References}
\addcontentsline{toc}{chapter}{Bibliography}
\bibliography{Bib}{}
\bibliographystyle{unsrt}


\end{document}